\documentclass[oldversion,letter]{aa} % for the letters and with the traditional abstract
\usepackage{natbib}
\usepackage{graphicx}
\usepackage{txfonts}
\usepackage{calc}
\usepackage{color}
\usepackage{rotating}
\usepackage[USenglish]{babel} % amerikanische Silbentrennung

\begin{document}

\title{Fine structures in the atmosphere above a sunspot umbra}

\author{L. Bharti\inst{1} \and J. Hirzberger\inst{1} \and S. ~K. Solanki\inst{1,2}}
\institute{Max-Planck Institut f\"{u}r Sonnensystemforschung, 37191 Katlenburg-Lindau, Germany \and School of Space Research, Kyung Hee University, Yongin, Gyeonggi, 446-701, Korea}
\offprints{L. Bharti \email{bharti@mps.mpg.de}}

\date{Received ...; accepted ...}

\abstract{We present simultaneous photospheric and chromospheric observations of
the trailing sunspot in NOAA 10904, obtained with the Swedish Solar
Telescope (SST) La Palma, Canary Islands. Time series of high resolution
\ion{Ca}{ii}\,$H$ images show transient jet-like structures in
sunspot umbrae are elongated, which we call umbral microjets. These jets are directed roughly
  parallel to nearby penumbral microjets, suggesting that they are aligned with the background magnetic field. In general, first a
bright dot-like structure appears, from which a jet later emerges, although some
jets appear without an associated chromospheric dot. Bright photospheric umbral dots are associated with umbral
microjets arising in the outer umbra. Nevertheless, a one-to-one correspondence between
jet-like events and underlying umbral dots is not seen. They are typically less than 1\arcsec  ~long and less than 0\farcs3 wide. The typical lifetime of umbral microjets is around one minute. The brightness of these structures increases from the center of the umbra towards the umbra-penumbra boundary along with the brightness of the local background.}


\titlerunning{Umbral fine structures}
\authorrunning{L. Bharti \and J. Hirzberger \and S.~K. Solanki}

\maketitle

\section{Introduction}
%Sunspots remain intriguing and partly enigmatic objects that have
%escaped a consistent and complete description for many years. Our
%understanding of the structure and physical processes acting in the
%umbra and the penumbra have recently been significantly advanced by
%advanced instruments including the Swedish Solar Telescope (Scharmer et al. 2003),
%Hinode (Tsuneta et al. 2008) spacecraft and numerical simulations (Sch\"ussler \& V\"ogler 2006,
%Heinemann et al. 2007, Scharmer et al. 2008, Rempel et al. 2009a, b).
%In particular, there is now strong evidence for the existence of
%magnetoconvective processes in the umbra and penumbra. Sunspot fine structure
%seen in photosphere i.e. umbral dots, light bridges and perumbral filaments
%are assumed to be due to overturning convection in the presence of a vertical
%and horizontal magnetic field respectively.
%However, observations of sunspots in chromosphere shows a different structure.

Fine-scale jet-like features are observed at the edges of penumbral filaments and referred
to as penumbral microjets \citep{katusk07}. They represent one of the most exciting and enigmatic additions to sunspot dynamic features.
These jets are associated with transient brightening. They are aligned with the
the background magnetic field \citep{jurcek08}. So far, observations of these jets have been restricted to the penumbra.

The chromosphere above sunspot umbrae shows other dynamic phenomena, mainly umbral flashes, bright patches appearing with a 3-min periodicity
. They are thought to be the chromospheric counterparts of
photospheric umbral oscillations that steepen with height
due to the lower density and produce hot shocks  \citep{kneer81,turova83,socas00,ariset01,tiziot02,rouppe03,nagash07,tiziot07,socas09}. It has been suggested that they have the same origin as running penumbral waves, ring-like disturbances moving from the edge of
the umbra towards the penumbra \citep{tsirop00,rouppe03,bloom07}.
%In H$\alpha$ spectral line penumbral
%filaments extend radially outward from sunspot and forms super penumbra. The
%Evershed flow seen in photosphere show opposite direction in this line and
%known as reversed Evershed flow (Foukal 1971, Maltby 1975, Alissandrakis et al. 1988) .
Recent observations in the \ion{Ca}{ii}\,$H$ line obtained with Hinode
show a fine filamentary structure in umbral flashes having a similar size-scale as the penumbral jets \citep{socas09}. This encourages us to look at the umbral chromosphere with even higher resolution data for possible counterparts of penumbral jets. Further motivation for such a study is provided by the underlying similarity (at photospheric layers) between umbral dots and penumbral filaments revealed by stat-of-the-art MHD simulations \citep[e.g.][]{schussler06,rempel09a}. According to these simulations the difference in appearance is driven basically by the difference in inclination of the magnetic field: both features are manifestations of overturning convection. One idea for the origin of penumbral jets is that they are produced by magnetic reconnection between field twisted by roll-like convective motions and more laminar field lines \citep{magara10}. Similar process could also take place at the edges of umbral dots.

% This suggest
%that transient events depends on the filed topology of sunspot umbra and penumbra. Since sunspot features in
%the photosphere are tracers of the field geometry
%(i.e. penumbral filaments and peripheral umbral dots in inclined filed and umbral dots in
%vertical filed), it is worth to compare photospheric and chromospheric umbral
%features.

We use high resolution \ion{Ca}{ii}\,$H$ observations obtained with the Swedish Solar Telescope
to look for chromospheric jets in a sunspot umbra.
%This letter is organized as follows: in Section 2 we describe the observations, in Section 3 we describe the
%jet-like events in the umbra and their association with underlying photospheric umbral features. We
%present our conclusions in Section 4.

\section{Observations}

Multi-wavelength observations of the trailing sunspot of the active
region NOAA 10904 were obtained at the Swedish Solar Telescope (SST)
La Palma, Canary Islands
on August 13, 2006. The center of the field of view was located at a
heliocentric angle of $\theta= 40.15^\circ$ ($\mu = 0.76$), bf which corresponds to solar disk coordinates
x = −556\arcsec and y = −254\arcsec.

The sunlight was divided into a blue and a red channel close to the
f/47 focus of the SST by using a dichroic mirror plate.
The blue beam was further divided, passed through various interference filters and imaged on four Kodak
Megaplus 1.6 CCD cameras. One of these cameras was fed with light from the
bandhead of the CH molecule at $\lambda$ = 4305$\AA$ (6.5\,\AA~  FWHM,
G-Band) formed in the photosphere. Two
cameras, one of them slightly out of focus (for phase-diversity
wavefront sensing), were fed with the light passing an
11\,\AA\ interference filter with a central wavelength of 4363\,\AA
~(G-continuum). In front of the 4th camera a tiltable narrow-band
\ion{Ca}{ii}\,$H$ filter (FWHM = 1.1\,\AA ) sampled the line at two wavelength positions:
(1) in the line center and (2) in the blue wing at 0.6 \AA\ from the line center, probing
 the lower chromosphere and mid-upper photosphere respectively.
%The calcium Ca II H 3968.5A line
%was recorded at two wavelength points by tilting this filter, once in the line centre
%and once in the blue wing, approximately 0.6A out of the
%line centre.
For these two wavelength positions, four frames at a time were recorded alternatingly.
The frame rate was between 3 and 4 frames per second (G-continuum) and the exposure time was 13\,ms for all four cameras in
the blue beam. The size of one pixel corresponds to
0\farcs041 in the blue beam.

The G-continuum and the \ion{Ca}{ii}\,$H$ data were reconstructed by
means of speckle interferometric techniques \citep[see][who analysed the same observations, for more detail on the instrumental setup and image
reconstruction]{hirzberg09}. After reconstruction, a
cadence of 19\,s was achieved for these time series. For the present
analysis we used only time series of G-continuum and \ion{Ca}{ii}\,$H$
images, each of them containing 151 images.

Since umbral brightness varies from the dark nuclei toward the penumbral boundary and bright patches are often dominated by the highly variable umbral flashes,
we selected the darkest parts of the umbra as the intensity reference used to normalize intensity in each image.
 To isolate bright features in the umbra, an intensity threshold of 5$\sigma$ of the mean local
background was applied and a binary mask created, with all pixels below the threshold
being set to zero and those above to 1. The selected threshold thus depends on the location of the jet in the umbra.
The local background was selected by choosing a box (size twice the dimensions of feature) with the bright feature at its center. The mean intensity of the box excluding the bright feature is defined as the local background.
We use mathematical morphological operations \citep[MMO,][]{gonz08} on binary images to enhance and extract the features of interest.
Thus obtained binary images are then ``opened'' \citep{haralick87} with a 3
pixel by 3 pixel square-shaped operator, using  ``dilation''
and ``erosion'' operations \cite[see][for more detail]{maurya10}. The width of the jet is defined as the width measured at the middle point of the length of the segmented jet.
%Figure~1 shows an example
%of the resulting segmentation of a typical umbral bright structure.

\section{Analysis}

\begin{figure}
\vspace{9mm}
\centering
\includegraphics[width=90mm,angle=0]{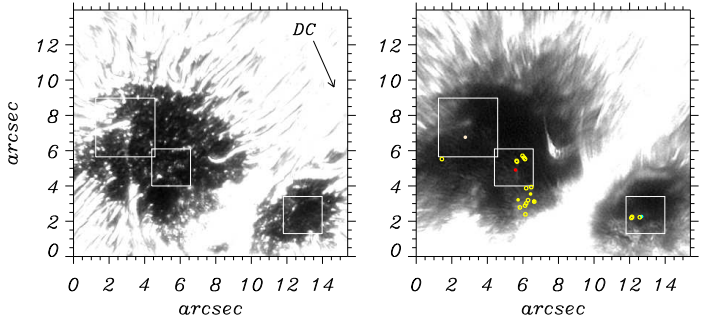}
\vspace{-5mm}
\caption{Left panel: the observed sunspot imaged in G-continuum.
The grey-scale has been chosen to better show umbral features.
The direction towards disk center is indicated by the arrow
  labeled 'DC' at the upper right corner; right panel: simultaneous
  and cospatial \ion{Ca}{ii}\,$H$ line core image of the same
  sunspot. The grey-scale is chosen to enhance fine structures
  in the umbral chromosphere. Circles show the locations of all
  jet-like events identified in the time series. Circles filled in
  different colors mark the locations of events discussed in the main
  text. White boxes show the fields-of-view  of events discussed in the text and shown in Figs. 2-4 (from left to right).
  The temporal evolution is shown in a movie available in the on-line edition.
  The movie uses a slightly different grey-scale to highlight the features.}
\end{figure}

Figure 1 shows simultaneous images of the sunspot in the G-continuum
and in the \ion{Ca}{ii}\,$H$ line core. In both the images a gray-scale is
chosen to better show the umbral features. The
sunspot is divided into two umbrae by a broad lightbridge. In the
larger umbra, besides umbral dots, a dark nucleus centered at
x=5\arcsec and y=3\arcsec is visible. Also,
penumbral filaments penetrate well into the larger umbra. The smaller
umbra shows peripheral umbral dots and a dark nucleus in its center.
 The penumbral intrusions visible in
the photosphere can also be identified in the \ion{Ca}{ii}\,$H$ image, but
with lower contrast. Central umbral dots are not visible in the umbral chromosphere. This
suggests that due to the narrow width of the interference filter used
in the observations, the photospheric contribution is efficiently blocked
in these images. In addition, in the chromosphere many jet-like bright
structures, located above penumbral filaments (penumbral microjets,
cf. \cite{katusk07}) as well as above the lightbridge
and several penumbral intrusions are visible.

\begin{figure}
\vspace{-18mm}
\hspace{8mm}
%\centering
\includegraphics[width=160mm,angle=0]{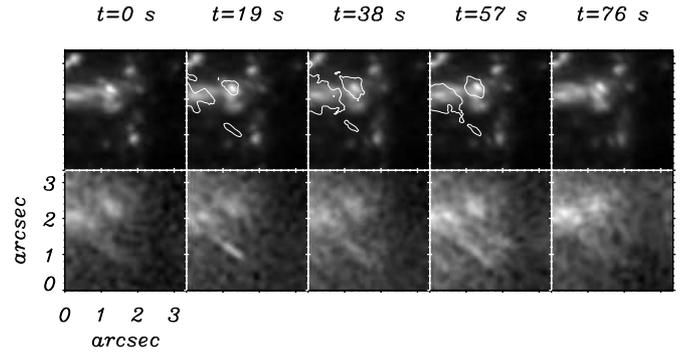}
\vspace{-3mm}
\caption{Upper row: time series of G-continuum images; lower row:
cotemporal and cospatial images in the \ion{Ca}{ii}\,$H$ line core
showing temporal evolution of a jet event located close to
penumbral filaments. The
  location of this event is marked by the almond colored circle at x=2\farcs75 and y=6\farcs75 in the
  right panel of Fig. 1.}
\end{figure}

\begin{figure}
\vspace{-18mm}
\hspace{5mm}
%\centering
\includegraphics[width=150mm,angle=0]{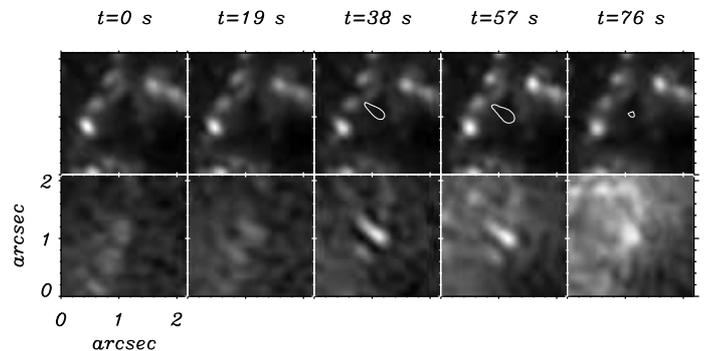}
\vspace{-8mm}
\caption{The same as Fig. 2 for a jet-like event located in the center of
  the larger umbra. The location of this event is marked by a red circle at x=5\farcs5 and y=5\farcs0 in
  the right panel of Fig. 1.}
\end{figure}

\begin{figure*}
\vspace{-17mm}
\centering
\includegraphics[width=165mm,angle=0]{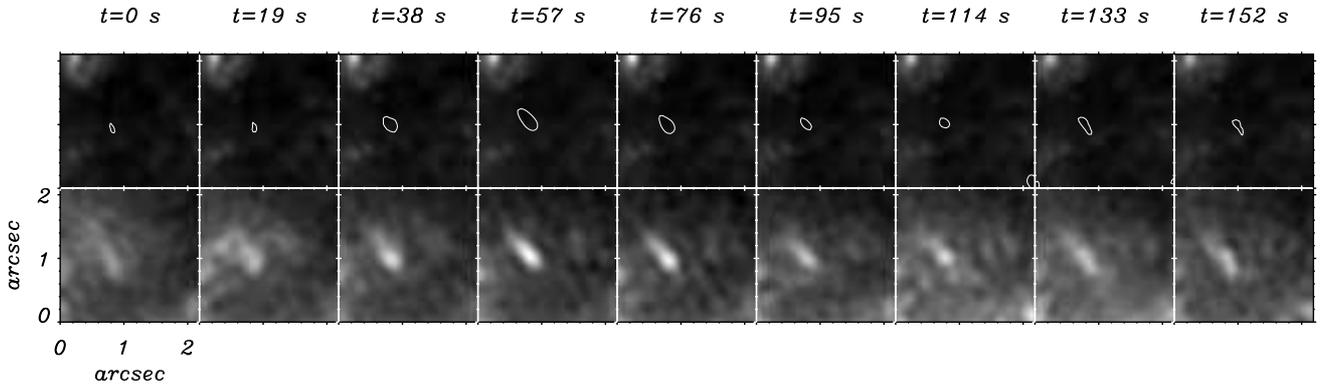}
\vspace{-7mm}
\caption{The same as Fig. 2 for a jet-like event in the
  central part of the smaller umbra. The location of this event is marked by the cyan
  circle at x=12\farcs60 and y=2\farcs25 in Fig. 1.}
\end{figure*}

An animation of the thresholded \ion{Ca}{ii}\,$H$ images (see online mpeg movie attached to Fig. 1) reveals  several
dynamic phenomena in the umbral chromosphere. The well-known umbral
flashes are visible throughout the umbra. They contain very fine
filamentary structures, confirming the findings of \cite{socas09}. Apart from the bright patches produced by umbral
flashes, several bright elongated jet-like structures can be
recognized. In most cases first a bright dot-like structure
appears and then an elongated jet-like feature becomes
discernible in subsequent images, although sometimes the jet
appears without the prior appearance of a dot-like structure. These jet-like structures are always
oriented in a direction parallel to nearby
penumbral microjets.  When an umbral flash passes through these
structures their brightness increases, but they are also clearly present outside umbral flashes. This implies that
these elongated structures are quiet distinct from the fine filamentary structure of umbral
flashes.

An example of the evolution of a jet-like event close to
the umbra-penumbra boundary is illustrated in Fig. 2.
In the \ion{Ca}{ii}\,$H$ images (lower row), a prominent bright elongated
structure becomes visible, starting at $t=19$\,s. No dot-like structure is seen
before the appearance of this jet.  Due to the location of the limb towards the left the jet is expected to be
expanding in the upper left direction although the expansion occurs too fast to be followed at our cadence. The jet's brightness decreases in the
subsequent frames, but due to the passage of an umbral flash, at $t=57$\,s (the diffuse brightening in the upper part of the frame) its brightness is again enhanced.  In the G-continuum images, shown in
the upper row, a row of bright peripheral umbral dots is visible. Assuming a
formation height difference of 500 km between these two spectral
regions and due to the inclined viewing angle, we should expect the
photospheric correspondent of this structure to be located along the
jet's major axis and shifted toward the lower right by 0\farcs82. From the \ion{Ca}{ii}\,$H$ intensity contours
overplotted on the G-continuum images the footpoint of
this jet is expected to lie close to the bright umbral dot at x=2\farcs0 and y=0\farcs7.

Figure 3 illustrates another example of the evolution of jets in the
\ion{Ca}{ii}\,$H$ images, now located in the central part of the umbra. In
the beginning, a faint dot is visible that develops an
elongated shape. At $t=38$\,s it is brightest and most strongly
elongated. From $t=76$\,s on it starts to becoming a round again and begins to fade. From $t=57$\,s to $t=76$\,s the brightness of the background
is enhanced due to the passage of an umbral flash. No clear relationship can be established between the supposed footpoint of the
jet and the chain of umbral dots visible in the corresponding G-continuum images.

%Two more examples are shown in Fig. 4 and 5 for chromospheric bright elongated jet like
%structure and corresponding photospheric umbral dots. Intensity contours are overplotted from Ca II H images on G-continuum
%images. Larger bright umbral dots in photospher are visible in Fig. 4.
%However, there is no corresponding underlying umbral dot along jet's major axis in chromospher. Similarly in Fig. 5, there are
%cluster of bright UDs in photosphere but there is no UD along jet's major axis that is seen in
%chromosphere.

%      Figure 6 illustrate jet like event in region of cluster of peripheral
%      UDs around x=10-12" and y=5-8". In 2nd and 3rd image there are two
%      bright dot like structure and then they coalesces and a elongated
%      feature visible in successive frames. Along major axis of this jet, in
%      photosphere cluster of bright UDs can be seen as suggestive from contours from Ca II H images.

%All 4 events have lifetimes of about 4\,min.

Figure~4 shows another example of the evolution of such an event,
this one located in the central part of the smaller umbra. No UD is present along the
 major axis of this \ion{Ca}{ii}\,$H$ jet in the G-continuum images.

In total, we have identified 19 and 4 events in the larger and the
smaller umbra, respectively. The 10 events identified in the
central region of both umbrae do not show a clear correlation with an
underlying photospheric umbral dot, whereas all 13 events identified close to the umbra-penumbra boundary occur close to UDs. In 2 of 13 events, a
photospheric UD is located close to where the footpoint of the chromospheric jet is expected to lie. The other 11
cases are observed over clusters of peripheral UDs so that it is difficult
to assign them to a single UD in the cluster.
%It is worth to noting that in all frames of the movie numerous large and bright UDs in
%the photosphere display no bright counterpart in chromospheric
%images (i.e. no contribution from photosphere due to narrow filter width).
%%It might be conjectured that there are two families of these jets: those that have
%%some association with underlying UDs and those that do not.
%%However, better statistics are required to establish this and if these families have different properties.
However, better statistics are required to establish if there is any correspondence between UDs and jet-like events in different
parts of sunspot umbra.

Histograms of various parameters associated with these events are
illustrated in Fig. 5. The histogram of normalized intensities with respect to the darkest part of the umbra,
integrated over the jet's areas for each snapshot in which they are present (solid line) shows a
broad range since their relative brightness depends on the location within
the umbra. The events identified in the central part of the umbra display a
lower brightness while events close to the umbra-penumbra boundary exhibit
higher brightness. This might be partially due to straylight. The histogram of maximum normalized intensity of each jet is overplotted (dashed line).

%Similarly, from the histogram of the jet's widths it might be
%guessed that there are two classes. One class with a mean width
%0\farcs2 belongs to long elongated jet-like structures. Another class
%belongs to jets that evolve from bright dot-like structures. In such cases, jet's end that close to dot is broad and
%another end of jet is narrow. Thus the jet's widths are larger as the
%widths measured in the middle point of there length which have a mean width of 0\farcs27.

The maximum width, of the umbral jets (Fig. 5b) is found to be 200 km which is half of the maximum width of penumbral microjets of about 400 km reported by \cite{katusk07}.
The lower widths found here may be due to the fact that the resolution of images presented here is higher, although umbral jets may also be narrower than their penumbral counterparts. This latter may well be the case since umbral jets are also significantly shorter than the penumbral ones (see panel c), where 1000-4000 km \citep{katusk07}. The histogram of the lengths of the structures shows a monotonic
decreasing trend towards longer structures (for lengths \textgreater 0\farcs4).
% The obtained lengths are in a range between 0\farcs2 and 1\farcs0.
The umbral-microjets are less long than penumbral microjets. Finally, the histogram of
lifetimes of 23 jet events shows that most of the jets live around 1\,min ( i.e. 3 frames of our observations). Jets found
in the smaller umbra show longer lifetimes of around 4\,min. These values are not too different from the life times of penumbral microjets reported to be up to two minutes by \cite{katusk07}.

Histogram of maximum local contrast for each jet is plotted in Fig. 6. The contrast of most umbral jet events is higher than the values reported by \cite{katusk07} (10 to 20 $\%$ with respect to underlying penumbral structure).

%\begin{figure*}
%\vspace{-15mm}
%\centering
%\includegraphics[width=175mm,angle=0]{figm22.ps}
%\vspace{-8mm}
%\caption{A sunspot image in blue continuum on 9 January 2007 taken by the SOT/BFI abord Hinode.
%Direction of disk center shown by arrow indicating 'DC' at upper right corner. 'A' shows location of
%time slice at head, middle section and between middle section and tail (from blow) of a penumbral filament. Location of time slices for penumbral
%filaments 'B', 'C','D' and 'E' at different position in filaments are marked. }
%\end{figure*}

%\begin{figure*}
%\vspace{-15mm}
%\centering
%\includegraphics[width=175mm,angle=0]{figm23.ps}
%\vspace{-8mm}
%\caption{A sunspot image in blue continuum on 9 January 2007 taken by the SOT/BFI abord Hinode.
%Direction of disk center shown by arrow indicating 'DC' at upper right corner. 'A' shows location of
%time slice at head, middle section and between middle section and tail (from blow) of a penumbral filament. Location of time slices for penumbral
%filaments 'B', 'C','D' and 'E' at different position in filaments are marked. }
%\end{figure*}

\begin{figure}
%\vspace{-45mm}
%\centering
\hspace{-2mm}
\includegraphics[width=\linewidth]{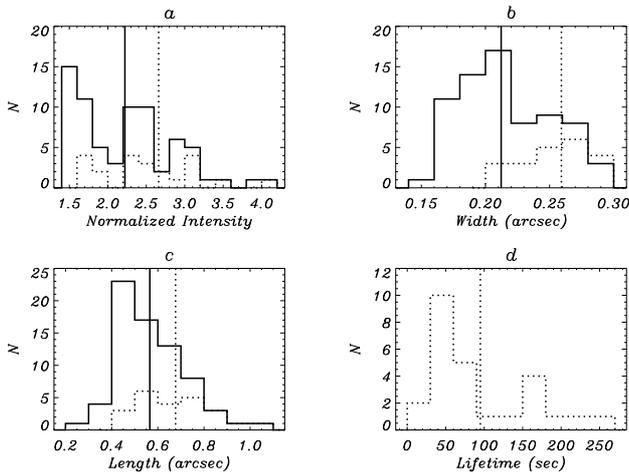}
\vspace{-4mm}
\caption{Histograms of various parameters of the identified jets. Solid lines: histogram of all values determined from
each snapshot (i.e. each jet counts multiple times), dashed lines: histogram of maximum values revealed by each jet (i.e. each jet is counted once). a:
  normalized intensity, b: width, c: length, d: lifetime. The vertical line indicates the
mean value.}
\end{figure}

\begin{figure}
%\vspace{-45mm}
%\centering
\hspace{-5mm}
\includegraphics[width=\linewidth]{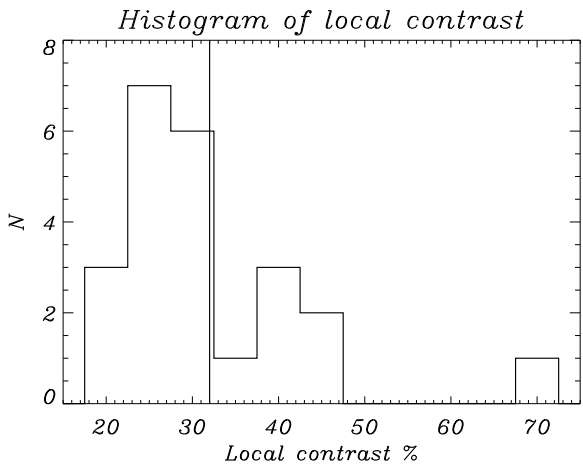}
\vspace{-4mm}
\caption{Histograms of maximum local contrast for each jet-like event. The vertical line indicates the
mean value.}
\end{figure}

\section{Discussion and conclusion}

We have presented the first high resolution observations of bright small-scale
elongated and short-lived
%structures of the umbral chromosphere,
structures in the atmosphere above a sunspot umbra, which we call
umbral microjets, in accordance with microjets observed by \cite{katusk07}.
These structures are elongated and in some cases appear as extensions of bright dot-like features.
%These elongated structures are inclined in the direction parallel nearby
%penumbral microjects which gives the impression of hot jet-like material in the umbra.
The brightness of these structures is higher than the brightness of penumbral microjets.

%Former studies of the chromosphere of sunspot umbrae by means of
From H$\alpha$ observations \cite{kitai86} and \cite{Kilcik2011} found
an association of photospheric umbral dots with chromospheric umbral dots.
 However, the width of the filter used for these observations was large enough (0.5\,\AA ~and 0.25\,\AA ~respectively) to allow
a photospheric contribution in the case of H$\alpha$~\citep{socas04} so that many of
these observed 'chromospheric' umbral dots may in fact have been photospheric
ones. However, they observed chromospheric umbral dots, but given the difference in the filter widths and line formation characteristics, it is not possible to directly relate their chromospheric umbral dots to the umbral jets found in the present study. The filter width used in the present study is narrow enough (1.1\,\AA~ for \ion{Ca}{ii}\,$H$) to
suppress the photospheric contribution, which is evident from the fact
that we cannot see photospheric umbral dots in the \ion{Ca}{ii}\,$H$
core data. Only larger UDs are seen in \ion{Ca}{ii}\,$H$ wing images obtained at 0.6 \AA\ from the line core and no jet like
events are visible in the \ion{Ca}{ii}\,$H$ wing images (not shown here). This indicates that the jet-like structures visible in
the \ion{Ca}{ii}\,$H$ core images, are chromospheric or temperature minimum region phenomena.
%It is also evident from comparison of cospatial photospheric and
%chromospheric images that even bright and larger UDs visible in
%G-continuum are not visible in Ca II H images.
Due to the formation height difference of the \ion{Ca}{ii}\,$H$ line
core images and the G continuum images, as well as the inclined
line of sight, it is difficult to find an accurate relationship with
photospheric umbral dots.
%angle and lot of observed dots in
%photosphere it is difficult to find accurate relationship with
%photospheric umbral dots with accuracy of size of UDs.
%%However, based on the evidence, we speculate that there are two families of
%%umbral micro-jets: those which are associated with
%%underlying photospheric UDs and those which are not.
We cannot rule out, however,
 that some jet-like events might be associated with tiny UDs
which are not resolved.

Simulated photospheric umbral dots
are associated with upflows in their central parts and with downflows
at their edges \citep{schussler06}.  In these simulations
jet-like outflows above the cusp are
seen.
%Below the cusp, high pressure builds up. This pressure is high enough to drive a
%material flow along the magnetic field lines above the cusp. The flow velocity of escaped material is
%about 1\,km\,s$^{-1}$. In such cases, the cusp acts like a
%safety valve as suggested by \cite{choudhuri86}. In the UDs simulated
%  by \cite{schussler06}, most of the upflowing matter
%turns over below the cusp after cooling and descends within the downflow
%channels due to overturning convection. A small part of matter escapes
%upward.
\cite{schussler06} proposed that ``outflows occur above most upflow plumes and
could possibly affect the chromospheric dynamics above sunspots''.  The simulations of \cite{schussler06} cannot predict effects to the chromosphere in
detail due to closed top boundary. These simulations only cover
the lower layers of the umbral atmosphere. Thus the observed umbral microjets may represent these
upflow jets that escape from the photosphere.

An alternative scenario is suggested by the simulations reported by \cite{bharti10}, who find that in larger UDs, strong downflows drag field lines
downwards and produce hairpin-like structures causing regions of
reversed magnetic polarity at the edges of umbral dots. Consequently the
umbral microjet may be caused by reconnection of opposite
polarity fields in the photosphere. A similar mechanism has also been proposed to explain penumbral
microjets \citep[see][]{katusk07,magara10}. However, reversed
polarity regions were found only in larger UDs with strong downflows
at the edges which could not explain those observed jets not found to be associated with UDs. Also, due to the limited spatial resolution of present
observations it is difficult to verify this scenario and to
demonstrate that reverse polarities really are present.
The magnetic field above the sunspot umbra might be quite complicated \citep{solanki03,socasnavarro2005,tritschler+etal2008}, which could provide alternate topologies and mechanisms for impulsive dissipation above sunpot umbrae.
%Moreover, Bharti et
%al. (2010) show that larger UDs reach higher into the umbral
%atmosphere.  In our study we do not find a correlation between these
%jet like events and the size of UDs.On the other hand,
%These events are also not associated with umbral flashes, so that they are unlikely to be caused by oscillations or
%wave phenomena.
%Magara (2010) presented
%a magnetrohydrodynamic model describing the production of penumbral
%microjets. Reconnection takes place between the highely twisted field of a horizontal penumbral flux
%tube with the background magnetic field that is
%directed in vertical direction. In this region magnetic field has a
%transitional configuration. A jet-like structure produce in this intermediate
%regian due to reconnection. The similar senerio may be possible for observed
%jet-like structure in umbra observed in present study.

%At this point no clear-cut mechanism can explain these jet like events
%in umbral chromosphere. To understand the fine structure of the upper
%umbral atmosphere high resolution polarimetric data in different
%atmospheric heights and at different disk positions are needed. A
%first step might become possible with, e.g., the CRISP instrument at
%the SST.  Extensions of realistic MHD simulations into the chromosphere, of
%sunspot umbrae would be extremely helpful.

\begin{acknowledgements}

The Swedish 1-m Solar telescope is operated on the island of La Palma by the
Institute for Solar Physics of the Royal Swedish Academy of Sciences in the
Spanish Observatorio del Roque de los Muchachos of the Instituto de
Astrof\'isica de Canarias. This work has been partly supported by the WCU grant No. R31-10016 funded by
the Korean Ministry of Education, Science and Technology. L.B. is grateful to the Inter University
Centre for Astronomy and Astrophysics (IUCAA) Reference
Center at the Department of Physics, Mohanlal Sukhadia University,
Udaipur, India, for providing computational facilities.

\end{acknowledgements}

%______________________________________________________________

\end{document}